# Reduced-Order Model of Power Converters to Optimize Power Hardware-In-the-Loop Technology in Dc-Distributed Systems

M. Sanz, D. Santamargarita, F. Huerta, D. Ochoa, A. Lázaro, A. Barrado
Universidad Carlos III de Madrid
Electronic Technology Department
Power Electronics Systems Group
Avda. Universidad, 30; 28911, Leganés, Madrid, SPAIN
Tel.: 34-1-6246024; FAX: 34-1-6249430
E-mail: marina.sanz@uc3m.es

*Abstract*— The Power Hardware-In-the-Loop (PHIL) technology provides a powerful tool for testing scenarios where there is a high-power interchange, in which the performance of field tests can be very complex or expensive. When performing PHIL simulations of systems with a high number of components, such as DC-distributed systems on a ship or aircraft, the use of switched or average models of the converters can require the use of expensive commercial real-time digital simulators (RTDS) reducing the advantages of these technology. This paper is focused on the proposal of a reduced order model of converters to be able to perform PHIL analysis of Dc-distributed systems using low resources of the required real-time digital simulator. The paper validates that the proposed reduced-order model is able to determine the stability on the Dc-distributed system in comparison with more complex converter models. Moreover, a comparison between both models regarding the required resources in the implementation in a commercial RTDS platform is performed to validate the benefits of the proposed model in performing PHIL analysis of large power distribution systems.

## I. INTRODUCTION

In the last decade, Power Hardware-In-the-Loop (PHIL) technology has emerged as a useful tool for the analysis and design of power systems. Based on the combination of simulated and physical systems, the PHIL technology offers a realistic, controlled and easily reconfigurable testbed that overcomes majority of techno-economical obstacles associated with field demonstrations and stand-alone hardware testing [1]. DC-distributed systems on an aircraft (Fig. 1) normally uses PHIL technology to test the behavior of a particular hardware equipment connected to a real time simulated system composed of several elements.

In a PHIL system (Fig. 2), a real-time digital simulator (RTDS) simulates part of a power system, also known as rest of the system (ROS), which is connected to a physical power hardware under test (HUT), by means of a power interface (PI). It is necessary to define an interface algorithm (IA) that determines how the exchange of signals between the physical and simulated elements will be. The PI and IA play a key role in the stability analysis of the power system by introducing additional dynamics and time delays [2], [3], [4].

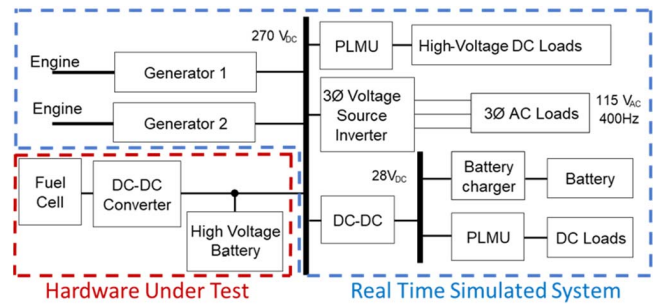

Fig. 1. PHIL system applied to distributed power system of an Electric Aircraft

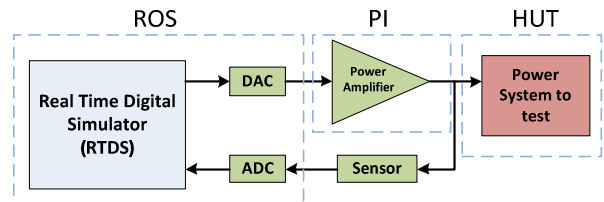

Fig. 2. PHIL system outline

This paper focuses on performing an optimization on the RTDS, using a reduced-order model of the power converters which can emulate the fundamental behavior of the power converters acting as a load in the system shown in Fig. 1. This optimization will lead to a decrease in the time-step and in the resources used in the RTDS, which means that larger and more complex systems can be analyzed with a better resolution in time, using less-expensive platforms.

The model to be presented is an adaptation of the reduced order input impedance model of switched converters depicted in [5]. The reduced order model is based on the behavior observed at the input port of a regulated converter in small-signal (Fig. 3). On one hand, the converter behaves as constant power load, due to the action of the control loop, that is represented as a negative resistor in small signal ($R_{CPL}$) at frequencies below the bandwidth of the converter and it is given by (1) where $V_i$ is the input voltage, $\eta$ is the efficiency and $P_o$ is the output power. On the other hand, at high frequency the effect of the input filter capacitor $C_i$ usually is predominant over the constant power load effect and the control loop effect, defining therefore the total input impedance of the converter $Z_{it}$ in (2).

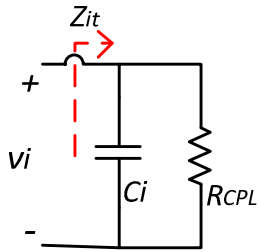

Fig. 3. Proposed model for small-signal input impedance

$$R_{cpl} = \frac{-V_i^2 \cdot \eta}{P_o} \qquad (1)$$

$$Z_{iT} = \frac{R_{cpl}}{1 + R_{cpl} \cdot C_i \cdot s} \qquad (2)$$

Previous publications have shown how the model correctly fits to the input impedance real value in most of the frequency range. Thanks to this it can be used in the stability analysis of the converter interconnection based on the study of the output and input impedances of each converter [6]. However, the reduced-order converter model represents only the behavior of the converter in small signal, and hence PHIL simulations cannot take advantages in order to analyze the power system stability.

In this paper, the proposed reduced-order model has been adapted to be applied to the stability analysis of a Dc-distributed system. A comparison between the implementation of the proposed reduced-order model in comparison with more complex converter models in a commercial RTDS is given to provide the advantages regarding the optimization of the resources.

This article is organized as follows. Section II presents the stability analysis used in PHIL simulations, Section III presents the proposed reduced-order model to be used in PHIL simulations and Section IV shows the validation of the model in the stability analysis of a Dc-distributed system in PHIL simulations.

## II. STABILITY ANALYSIS OF POWER HARDWARE-IN-THE-LOOP SYSTEMS

In order to study the stability in the connection of two individually stable hardware system, the connection can be modeled as the serial connection of the small signal output impedance of the source system $Z_S$ and the small signal input impedance of the load system $Z_L$ as it can be seen in the Fig.4. The stability in this interconnection is determined by the open loop expression (3).

In order to do the stability analysis in this paper the GMPM criterion [6] will be used. The GMPM criterion ensures that when the conditions of (4) are met the connection of the two systems will be stable, these conditions are equivalent to the Nyquist contour does not encircle the (-1, 0j) point as long as there are no poles on the right half plane.

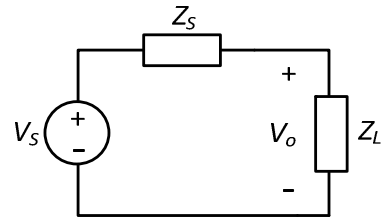

Fig. 4. Connection model of two systems based on their input and output impedances

$$T_{OpenLoop}(s) = \frac{Z_S(s)}{Z_L(s)} \qquad (3)$$

When $\|Z_S\| \ll \|Z_L\|$ must be satisfied that

$$|\arg(Z_S) - \arg(Z_L)| \leq 180° \qquad (4)$$

Focusing the stability study on a PHIL system, consisting of a real converter (Hardware Under Test, HUT) connected to a real time simulator that simulates part of a power system (Rest Of the System, ROS) via a power interface and using the interface algorithm known as Ideal Transformer Model (ITM) [2]. The connection diagram shown in Fig. 5 is obtained.

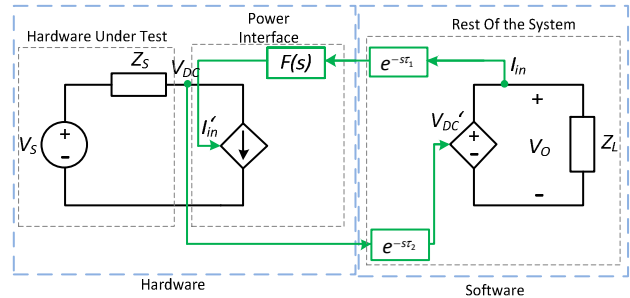

Fig. 5. PHIL connection using the ITM as an interface algorithm

It can be observed how the study of the stability of a PHIL system differs from the study of stability in hardware systems (Fig. 4). As a consequence, as it depicted in Fig. 6, the PHIL system introduces the effect of the controlled power interface $F(s)$. The system also introduces the effect of delays, being the delay produced from measurement to simulation $\tau_1$ and the delay produced from the simulated system to the power interface $\tau_2$, these delays are modelled as $e^{-s\tau}$.

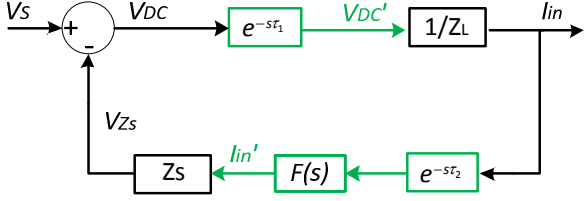

Fig. 6. Loop of PHIL system when using ITM

Hence, the open-loop system on which stability should be analyzed is shown in (5).

$$T_{OpenLoop}(s) = \frac{Z_s(s) \cdot F(s) \cdot e^{-s(\tau_1+\tau_2)}}{Z_L(s)} \quad (5)$$

### III. PROPOSED REDUCED-ORDER MODEL OF CONVERTERS

Since the proposed reduced order model shown in Fig. 3 represents only the behavior of the converter in small signal, it cannot be used directly in PHIL simulations as the negative resistance causes that the model acts as a source and not as a load in the steady state.

The proposed solution to this problem is to separate the voltage measured at the input in its small signal and in its steady state, once the behavior in small signal has been separated from the behavior in the steady state each voltage obtained will be used for the corresponding model. As it can be seen in Fig.7 the small signal and the steady state have been separated by means of a low-pass filter (LPF). After the filter, the steady state signal is obtained and if this value is subtracted from the original signal, the small signal value will be obtained. Each of these values will be used as a reference in the dependent voltage source corresponding to each model.

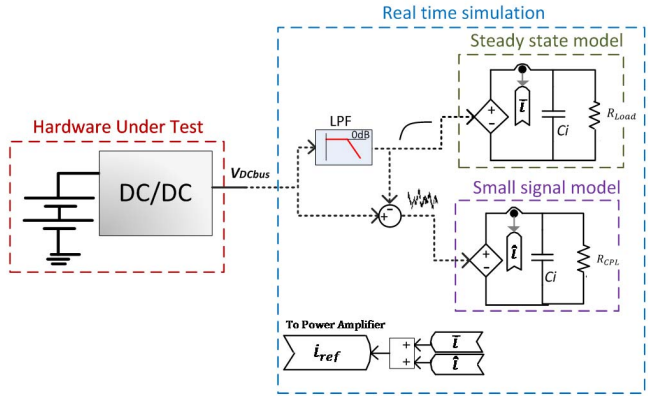

Fig. 7. Proposed reduced-order model for use in PHIL simulations

The steady state model represented by a positive resistor $R_{Load}$ is responsible for establishing the power consumption of the simulated converter that is calculated according to (6), where $P_o$ is the power consumption at the output of the converter and $\eta$ is the efficiency of the converter.

$$R_{Load} = \frac{V_{DCbus}^2 \cdot \eta}{P_o} \quad (6)$$

However, this steady state model does not reflect the stability of the system. To solve this problem and reflect the stability or the instability the small signal model previously shown in Fig. 3 has been used. Hence, another resistor $R_{CPL}$ with negative value according to (1) is used.

Finally, in order to obtain the input current reference that will be introduced into the power amplifier, the small-signal current $\hat{\imath}$ and the steady state current $\bar{\imath}$ must be added together, therefore the steady state and the small-signal will be taken into account.

### IV. VALIDATION

In order to verify the accuracy of the proposed reduced-order converter model for stability prediction in PHIL simulations, several tests has been performed.

#### A. Test 1: LC-filter and converter

For better understanding, a simplified power system has been considered for the validation (Fig. 8). An LC filter is used as a hardware device instead of a DC/DC converter and only one inverter will be used in the simulated part. The LC filter has been connected at the input of a two-level three-phase (2L-3Ph) grid-tied inverter with an input capacitor filter (Ci) of 1 µF, an output inductance filter of 510 µH. A DQ current control (PI type controller) has been designed [7] in order the inverter operates stable in stand-alone conditions for an output active power between 40 kW and 80 kW.

The parameters depicted in Table. I have been used for all the tests to be carried out in this section. Regarding PHIL system parameters, typical values [8] have been considered: $\tau_1$= 5 µs, $\tau_2$= 5 µs, being 15 kHz the bandwidth of the controlled power interface F(s).

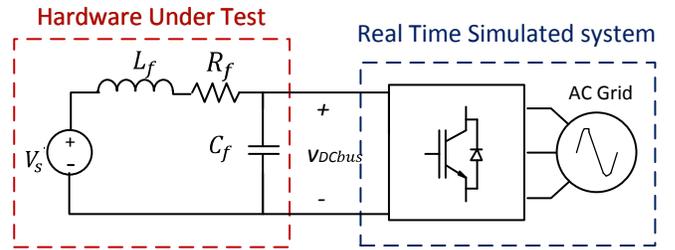

Fig. 8. Test 1: Block diagram of the considered power system

TABLE I. POWER SYSTEM SPECIFICATIONS

| Parameter | Value |
|---|---|
| Input Voltage ($V_s$) | 680 V |
| Line-to-Line output voltage ($V_{ac}$) | 380 $V_{rms}$ |
| Filter inductance ($L_f$) | 100 µH |
| Filter capacitance ($C_f$) | 0.1 mF |
| Filter resistance ($R_f$) | 0.1 Ω |
| Input inverter capacitance ($C_i$) | 1 µF |
| Output inverter inductance ($L_o/R_o$) | 510 µH/0.07 Ω |
| Inverter efficiency ($\eta$) | 100% |
| Measurement to simulation delay ($\tau_1$, tau1) | 5 µs |
| Simulation to power interface delay ($\tau_2$, tau2) | 5 µs |
| controlled power interface bandwidth (F(s), Fs) | 15 kHz |

In order to analyze the stability of the system in the considered operating conditions, the proposed reduced-order model has been considered. Stability analysis has been carried out using the GMPM criterion on the open-loop transfer function considering the PHIL effects $T_{OpenLoop}$ given by (5). As it can be seen in Fig. 9, bode and Nyquist diagram indicated that the system is stable when the inverter demands an active power of 40 kW and the system is unstable when the inverter demands 80 kW.

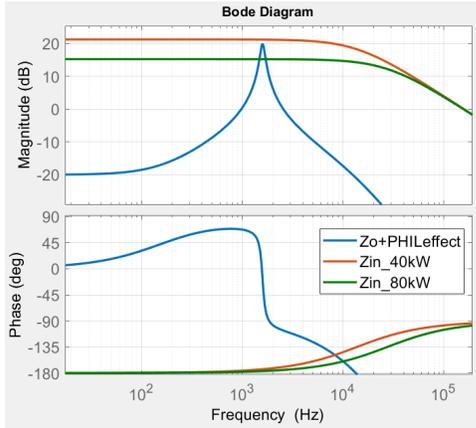

a. Bode diagram of the output (Zo) and input impedance (Zin)

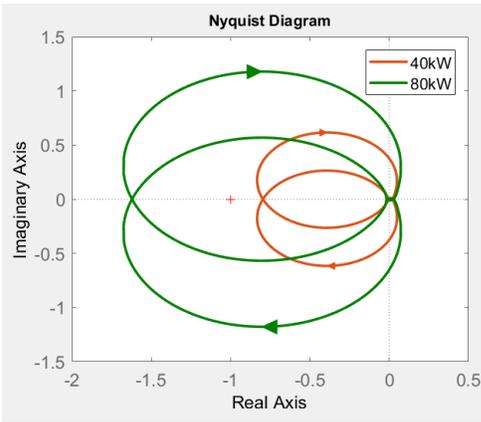

b. Nyquist diagram of the open loop ($T_{OpenLoop}$)

Fig. 9. Stability of the PHIL simulation of the considered power system

In order to validate the accuracy of the proposed reduced-order model regarding stability of the power system in PHIL simulations, the transient response of the whole power system had been simulated in the commercial software PSIM®, using the connections of the ITM interface algorithm, taking into account the corresponding delays and the controlled power interface. On one hand, an average model of the inverter has been considered in order to take into account the fundamental dynamics of the system in small-signal. On the other hand the proposed reduced-order model has been used. According to section III, in the proposed reduced-order model, the value of $R_{load}$ varies between 11.56 Ω and 5.78 Ω for 40 kW and 80 kW, respectively. In addition, the value of $R_{cpl}$ varies between -11.56 Ω and -5.78 Ω for 40 kW and 80 kW, respectively. The cut-off frequency of the LPF block has been chosen equal to 5 Hz for separation of the small-signal behavior of the selected inverter.

The considered systems are shown in Fig. 10 and Fig. 11, respectively and the transient response is shown in Fig. 12 and Fig. 13.

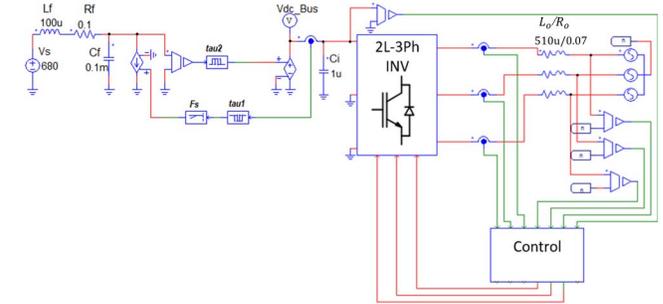

Fig. 10. PHIL simulation using a Two-Level Three-phase inverter

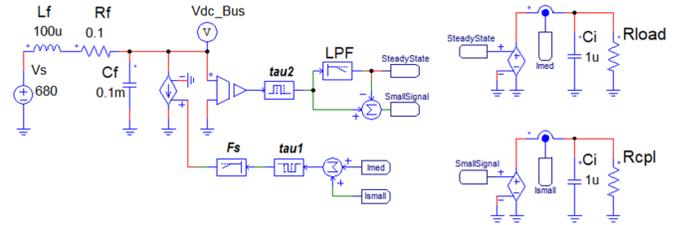

Fig. 11. PHIL simulation using the proposed reduced-order model

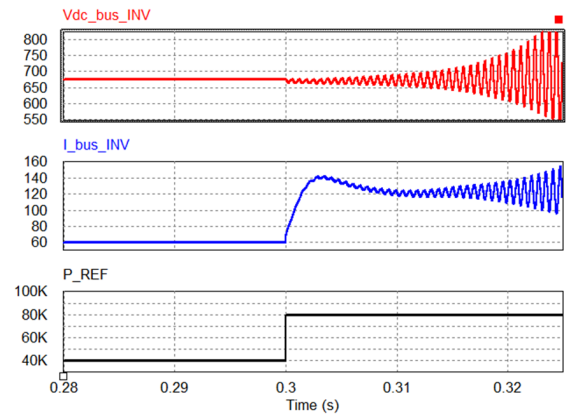

Figure 12. Voltages and currents on the simulated distribution bus obtained using a switched model

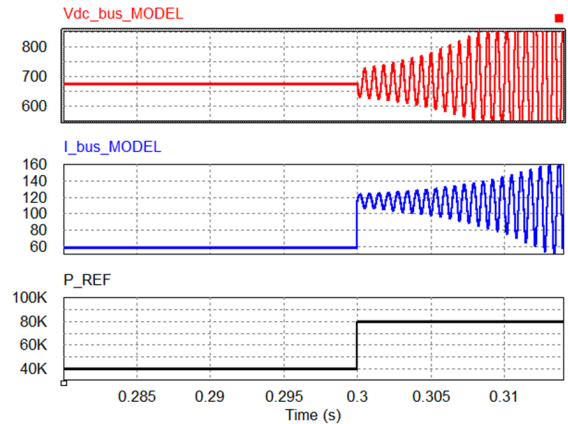

Figure 13. Voltages and currents in the simulated DC bus, obtained with the proposed reduced-order model

Fig. 12 and Fig. 13 shows the current (I_bus) and voltage (Vdc_bus) in the DC bus when a positive step of the active power reference (P_REF) is produced. As seen, the proposed reduced-order model is able to predict the instability of the system.

## B. Test 2: Multiple converters

Once the proof of concept has been presented, the accuracy of the proposed reduced-order model regarding has been validated in a more complex distribution system, where the source system (the LC filter) is changed by a DC-DC boost converter. In addition, in order to check that the model behaves correctly for multiple connections, another inverter will be added operating as a load, Fig. 14.

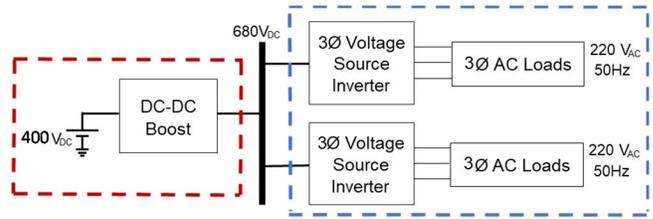

Fig. 14. Test 2. Block diagram of the considered system

Fig. 15 shows the schema of the connections used in simulation when the switched model of a Two-Level Three-phase inverter is used. The Fig. 16 shows the connections for the simulation with the proposed reduced-order model.

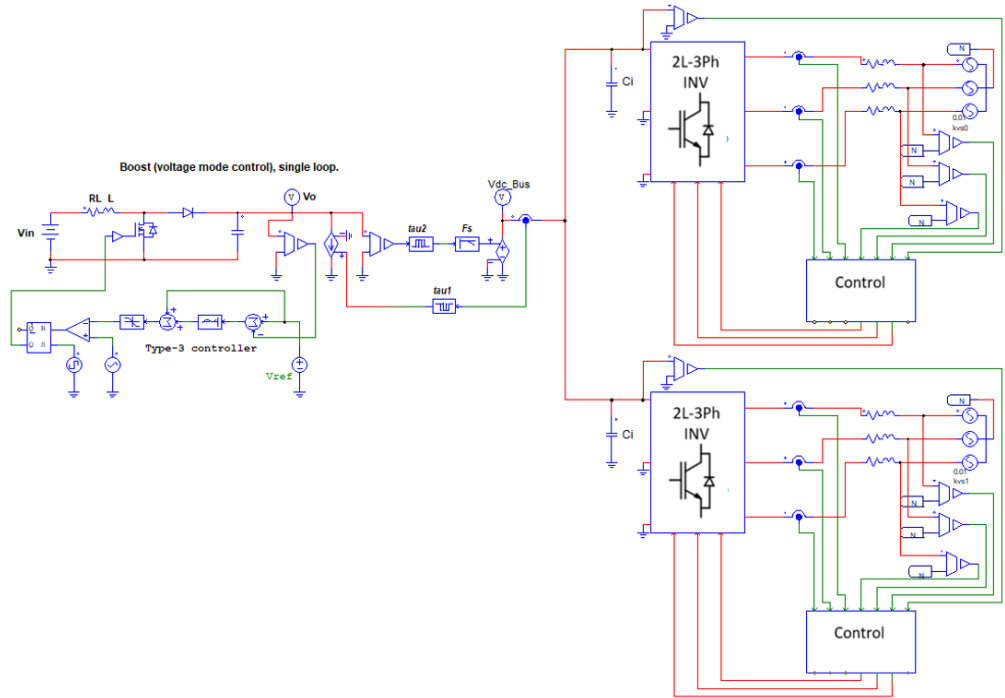

Fig. 15. Simulation of the PHIL interconnection of a Boost converter to two Inverters acting as a source.

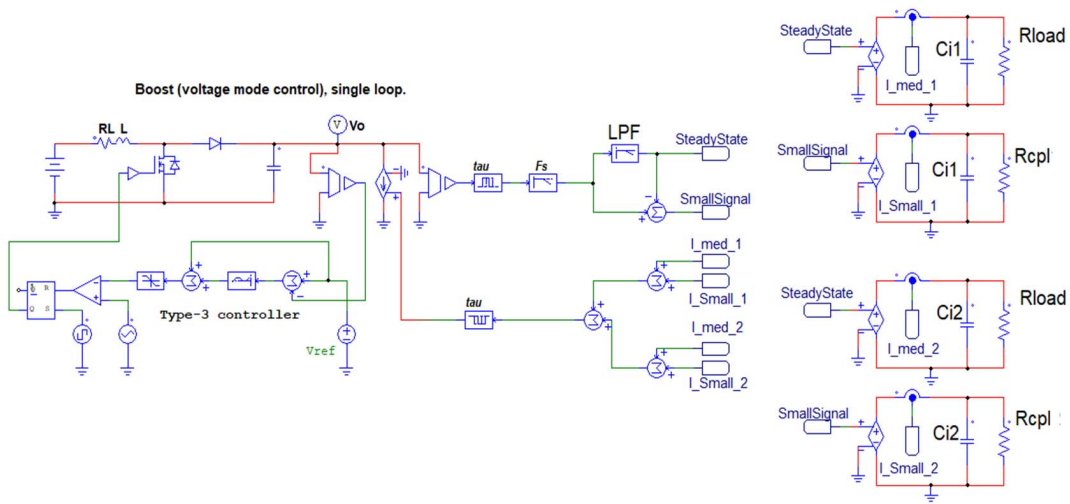

Fig. 16. Simulation of the interconnection using the proposed reduced order model

The parameters of the considered regulated DC-DC Boost converter are summarized in Table II.

TABLE II. DC-DC CONVERTER PARAMETERS

| Parameter | Value |
|---|---|
| Switching frequency | 200 kHz |
| Input converter Inductance (L/RL) | 50 µH/1 nΩ |
| Output inverter Capacitance ($C_o/R_o$) | 47 µH/10 mΩ |

Both inverters are equal and their parameters and the PHIL simulation parameters are the same that the ones described for Test 1. Regarding the reduced-order model, the values of $R_{load}$ and $R_{cpl}$ has been calculated according to (6) and (1), respectively, taking into account the power levels of the transients specified in Fig. 17 and Fig. 19 for each inverter (Inv1 and Inv2).

As it can be seen in the Fig. 17 and in Fig. 18 the proposed reduced-order model is able to simulate the consumption of the converter, as well as the instability of the DC bus when it is produced.

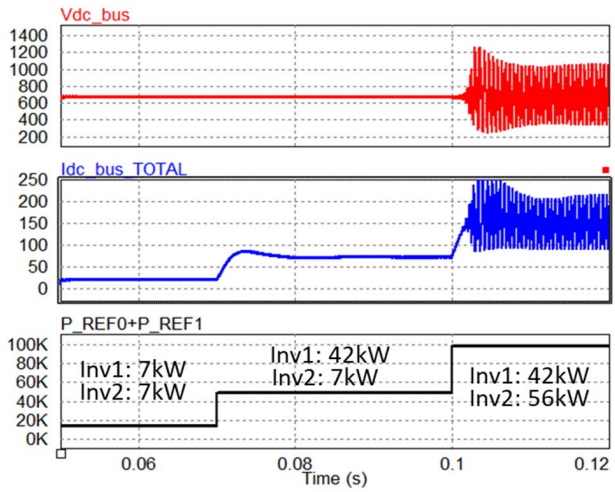

Fig. 17. Voltages and currents on the simulated distribution bus obtained using the model of Two-Level Three-phase inverter

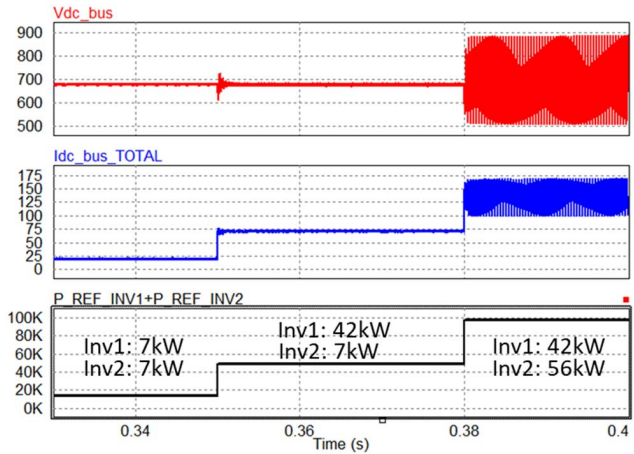

Fig. 18. Voltages and currents in the simulated DC bus, obtained with the proposed reduced-order model

### C. Test 3: Simulation in a real time digital system

In addition, the resources used by the commercial RTDS system *OP-4510* required to implement the PHIL system has been analyzed.

In order to show the potential benefits of the proposed reduced-order model in PHIL system, the previous Dc-distributed system consisting of five 2L-3Ph inverters has been considered. In order to show more clearly the advantages, the averaged model of the inverter has been considered instead of the switched model. Hence, an execution cycle time of 100 µs has been chosen. The average model is the one provided by the commercial simulation tool used by the RTDS system MATLAB ®.

As it can be seen in Table III, the RTDS system requires 2.27% of the execution cycle time (2.27 µs) when using the averaged converter model, while the RTDS system only requires 0.57% of the execution cycle time (560 ns) when using the proposed reduced-order model. Hence, the proposed reduced-order model of the inverter allows reducing about 4 times the execution cycle time in comparison with its average model.

TABLE III. RTDS RESOURCES OF THE DC-DISTRIBUTED SYSTEM ANALYSIS WITH FIVE 2L-3PH INVERTERS

| Average converter model | | Proposed reduced-order model | |
|---|---|---|---|
| Probes | Usage [%] | Probes | Usage [%] |
| Inversor Ts=1.0E-4[s] | 2,27% | ModOrdenRed...=1.0E-4[s] | 0,56% |
| Time steps [s] | | Time steps [s] | |
| sm_computat...8752E-5[s] | 2,27% | sm_computat...=1.0E-4[s] | 0,56% |
| New data acquisition | 0,04% | New data acquisition | 0,03% |
| Major comp...tion time | 1,21% | Major comp...tion time | 0,07% |
| Minor comp...tion time | 0,59% | Minor comp...tion time | 0,06% |
| Execution cycle | 2,27% | Execution cycle | 0,56% |

Hence, the proposed model provides the capability of analyzing the stability of Dc-distributed systems reducing significantly the requirements of RTDS system resources. Therefore, the proposed model allows analyzing more-complex systems or the use of less-expensive RTDS systems.

### D. Test 4: Experimental validation

Finally, the small-signal input impedance of the proposed reduced-order model has been experimentally validated using a three-phase inverter prototype of 30 W being the input Dc voltage 25 V and the input capacitor ($C_i$) equal to 10 µF.

A resistive load of 10 Ω per phase and an output inductive filter of 330 µH (MURATA 60B334C) per phase have been considered. Hence, the considered output power for the test is 4 W. The switching frequency is 100 kHz and a DQ control with a PI controller has been to obtain a proper performance of the converter in closed-loop conditions as shown in Fig. 19.

As seen in Fig. 20, the proposed reduced-order model of the inverter (Fig. 5) with Rload equal to 156.25 Ω and RCPL equal to -156.25 Ω provides an accurate estimation of the measured small-signal input impedance of the inverter for the PHIL system stability analysis.

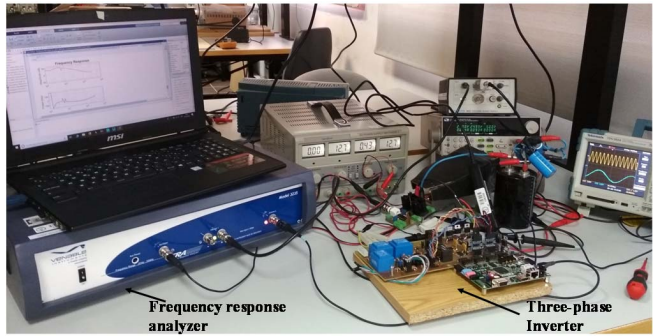

Fig. 19. Inverter prototype

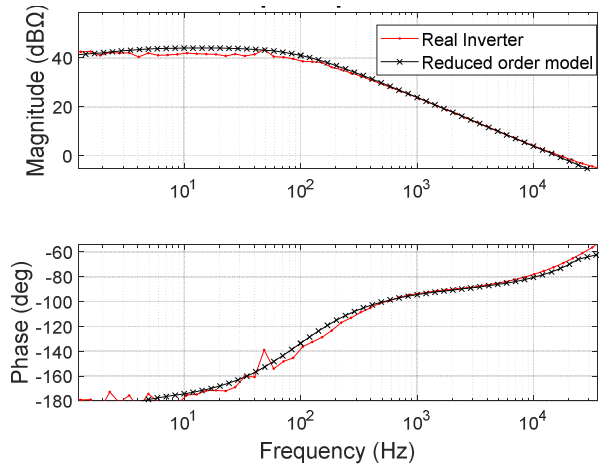

Fig. 20. Comparison of the model (black line) and measurements (red line). Small-signal input impedance of the inverter prototype for 4 W output power

## V. CONCLUSIONS

In this paper the use of a reduced order model has been proposed and validated to be used in the stability analysis of Dc-distributed systems by means of Power Hardware-In-the-Loop (PHIL) technology.

Moreover, a Dc-distributed system with five inverters as Dc bus load has provided a significant reduction of the execution cycle time (about 4 times) in the PHIL system when using the proposed model in comparison with its average model.

As a result, the proposed model allows analyzing more-complex systems with the same RTDS resources or using less-expensive RTDS systems.

### ACKNOWLEDGMENTS

This work has been partially supported by the European Regional Development Fund, the Ministry of Science, Innovation and Universities and the State Research Agency through the research project "Modelling and control strategies for the stabilization of the interconnection of power electronic converters" (FEDER/Ministerio de Ciencia, Innovación y Universidades – Agencia Estatal de Investigación/ _Proyecto CONEXPOT-2 (DPI2017-84572-C2-2-R)).